\documentclass[11pt]{cernrep}
\usepackage{graphicx}
\usepackage{cite,./mcite}
\newcommand{\pom}{I\!\! P}
\include{00README.XXX}

\begin{document}

\title{Exclusive Central Production and Diffractive W/Z Results from CDF~II\\
{\small\bf{\em HERA and the LHC, A workshop on the implications of HERA for LHC physics, 26 - 30 May 2008, CERN }}
}
\author{Konstantin Goulianos$^{1}$ and James L. Pinfold$^{2}$ 
\bigskip}
\institute{$^{1}$The Rockefeller University \\
1230 York Avenue, New York, NY 10065, USA\\ 
\medskip
$^{2}$The University of Alberta \\
Centre for Particle Physics, Edmonton, Alberta T6G2N6, CANADA }
\maketitle
\begin{abstract}
We report recently published  results on central exclusive production of di-jets and di-photons, and exclusive
 QED production  of $e^{+}e^{-}$ pairs. In addition,  we discuss preliminary results on exclusive  
 photoproduction of charmonium and bottomonium, exclusive QED production of $\mu^{+}\mu^{-}$ pairs, 
 and single diffractive W/Z production.  All the presented results were extracted 
 from data collected by the CDF~II detector  from $p\bar{p}$ collisions  at $\sqrt{s}$=1.96 TeV. 
 The implications of these results for the Large Hadron Collider (LHC) are briefly examined.
\end{abstract}

\section{Introduction}
We present results obtained by CDF~II at the Tevatron~\footnote{The presented results are from the CDF diffractive and exclusive physics  program of Run~II. This program relies on a system of special forward detectors, which include:  a Roman Pot  
Spectrometer (RPS) equipped with scintillation counters and a fiber tracker to detect and measure the angle and momentum of leading anti-protons,  a system of Beam Shower Counters (BSCs) \cite{BSC} 
covering the pseudorapidity  range 5.5 $ < |\eta| < $ 7.5 used to select diffractive events 
by identifying forward rapidity gaps and reducing non-diffractive background on the trigger level,  and  
two very forward (3.5 $ <~|\eta|~< $ 5.1) MiniPlug (MP) calorimeters~\cite{MP}, designed to measure energy and lateral position of both 
electromagnetic and hadronic showers. 
The ability to measure the event energy flow in the very forward rapidity region is vital for the
 identification of diffractive events in the high luminosity environment of Run~II.} in two broad areas: inclusive
diffraction and exclusive production. The main  goal of the Run~II inclusive diffractive program of CDF
 has been to understand  the QCD nature of the Pomeron
  \footnote{Diffractive 
reactions are characterized by the exchange of
 a spin 1 quark/gluon construct with the quantum numbers of the vacuum. In Regge theory, this exchange is the vacuum trajectory traditionally referred to as the Pomeron~($\pom$). Because the exchange is colorless, a large region in pseudorapidity
space is left empty of particles (this region is called a ``rapidity gap''). In perturbative QCD, 
the lowest order prototype of the Pomeron is the color neutral system of two gluons.} 
($\pom$) by measuring the diffractive structure function~\cite{FDJJ} $F^{D4}(Q^2,x_{Bj},\xi,t)$, where $\xi $ is the fractional momentum loss of the diffracted nucleon, for different diffractive 
production processes.  In addition, the possibility of a composite Pomeron is being investigated 
by studies of very forward jets with a rapidity gap between the jets. Important results  are
 the observation of a breakdown of QCD factorization in hard diffractive 
processes, expressed as a suppression by a factor
of $\cal{O}$(10) of the production cross section relative to theoretical expectations, and the  breakdown 
of Regge factorization in soft diffraction by a factor of the same magnitude \cite{FDJJ}. 
Combined, these two results support the hypothesis that the breakdown of factorization is 
due to a saturation of the rapidity gap formation probability  by an exchange of a color-neutral 
construct of the underlying parton distribution function (PDF) of the proton~\cite{lathuile}. Historically, such an exchange is 
referred to as the Pomeron. 
Renormalizing the ``gap probability'' to unity over all ($\xi$, t) phase space corrects for the
unphysical effect of overlapping diffractive rapidity gaps and leads to agreement between
theory and experiment (see \cite{lathuile}).

Central exclusive production in $p\bar{p}$ collisions is a process in which the $p$ and $\bar{p}$ remain
intact and an exclusive state X$_{excl}$ is centrally produced: $p + \bar{p} \rightarrow  p + X_{excl} + \bar{p}$.
The primary motivation for studying exclusive physics
at the Tevatron is to test the feasibility of using exclusive production to search
for and study the Higgs boson as well search for other new physics at the LHC \cite{KHOZEAGAIN}. 
In leading order QCD, exclusive production occurs through gluon-gluon fusion, while an 
additional soft gluon screens the color charge allowing the protons to remain intact~\cite{KHOZE2000}. This mechanism, historically termed 
Double Pomeron Exchange (DPE), 
 enables exclusive production of  di-jets~\cite{FDJJ}, $\gamma\gamma$ \cite{KHOZEGAMGAM},  and 
 the  $\chi^{\circ}_{c}$ \cite{KHOZECHIC} at the Tevatron, whereas at the LHC, where central
masses up to several hundred GeV are attainable, new central exclusive
channels open up, as for example W$^{+}$W$^{-}$ and Z$^{0}$Z$^{0}$.  
 While the main effort at the LHC  is directed toward
searches for inclusively produced Higgs bosons, an intense interest is  developing in exclusive
Higgs production, $p + \bar{p} \rightarrow  p + H + \bar{p}$. This production channel presents several
advantages, as for example the production of clean events in an environment of suppressed QCD background for the main Higgs decay mode of $H\rightarrow b_{jet}+\bar b_{jet}$ due to the $J_z=0$ selection rule~\cite{KHOZEAGAIN}.
Exclusive production can also occur through photoproduction 
($\pom$~-~$\gamma$ fusion), yielding charmonium and 
 bottomonium. The same tagging technique 
can also be utilized to select $\gamma p$, or  $\gamma q$ and $\gamma g$
interactions at the LHC, for which the energy reach and the effective luminosity are
 higher than for $\gamma\gamma$ interactions.

Additionally, exclusive production  of  central lepton pairs,  $\gamma\gamma \rightarrow l^{+}l^{-}$ ($l = e,\mu,\tau$),
via  two-photon exchange has been observed at CDF \cite{EXCLEPEM}.
Tagging two-photon production offers a significant extension of the LHC physics program.\footnote{The effective luminosity 
of high-energy $\gamma\gamma$ collisions reaches $\sim$1\% of the $pp$ 
luminosity, so that the standard detector techniques used for measuring very forward proton scattering 
should allow for a reliable extraction of $\gamma\gamma$ results.}  Particularly exciting 
is the possibility of detecting two-photon exclusive W$^{+}$W$^{-}$, Z$^{0}$Z$^{0}$, Higgs boson 
and new physics  production at the LHC \cite{TWOPHOT}. The deployment of
forward proton detectors at 200 m and 420 m (FP420 project) from the interaction point of ATLAS
and CMS, in order to exploit the above mentioned forward physics scenarios, is currently
under consideration \cite{FP420}. Two-photon exclusive production of 
lepton pairs will provide an excellent monitoring tool of the tagging efficiency and energy scale of the detectors of the FP420 project. 
These events can also  be used  for several
systematic studies, including luminosity normalization and contributions from 
inelastic production or accidental tagging.

\section{Central Exclusive Production}

\begin{figure}[htb]
\begin{center}
\includegraphics[width=24pc]{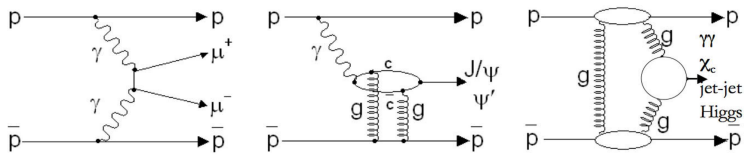}
\caption{Leading order diagrams for three types of exclusive process: $\gamma\gamma$ interactions (left), $\gamma\pom$  fusion or photoproduction (middle), and $gg$ $t$-channel color-singlet two-gluon exchange
(right). Higgs boson production proceeds via the  $gg$ diagram.}
\label{Fig:feynman}
\end{center}
\vglue -2em
\end{figure}

Exclusive production is hampered by expected low production rates \cite{KHOZEAGAIN}. As rate calculations are model
dependent and generally involve non-perturbative suppression factors, it is sensible to calibrate
them against processes involving the same suppression factors but have high enough  production
rates to  be measurable at the Tevatron. The leading order diagrams relating to the exclusive
central production processes discussed in this paper are summarized in Fig.~\ref{Fig:feynman}.

\subsection{ Exclusive Di-jet Production}

The process of exclusive di-jet production, which has been observed by CDF in Run~II data \cite{ PRDDIJET},
 proceeds through the same mechanism as $\gamma\gamma$, $\chi^{\circ}_{c}$, and Higgs production, as shown in Fig.~\ref{Fig:feynman}.
The analysis strategy  developed to search for exclusive di-jet production  
is based on measuring the di-jet mass fraction, R$_{jj}$ , defined as the 
di-jet invariant mass M$_{jj}$ divided by the total  mass of the central system: 
R$_{jj}$ = M$_{jj}$/M$_{X}$.\footnote{The mass M$_{X}$  is obtained from all calorimeter towers with energy above the 
thresholds used  to calculate $\xi^{X}_{\bar{p}}$, while M$_{jj}$ is calculated from calorimeter
tower energies inside jet cones of R=0.7, where R=$\sqrt{\Delta\phi^{2} + \Delta\eta^{2}}$. 
The exclusive signal is extracted by comparing the R$_{jj}$ distribution shapes of DPE di-jet data and simulated di-jet events obtained from a Monte Carlo (MC) simulation that does not contain exclusive di-jet production.}  The POMWIG MC \cite{POMWIG} generator along with a detector simulation are used to simulate the DPE di-jets.
The signal from exclusive di-jets is expected to appear at high values of R$_{jj}$, smeared by 
resolution and gluon radiation effects. Events from the inclusive DPE production process $p$ + $\bar{p}$ $\rightarrow$  $p$ + gap  +[ X + jj] + gap (the leading $p$ is not observed in CDF~II)
are expected to contribute to the entire R$_{jj}$ region. Any such events within the exclusive 
R$_{jj}$ range contribute to background and must be subtracted when evaluating exclusive production rates.

\begin{figure}[htb]
\begin{center}
\includegraphics[width=32pc]{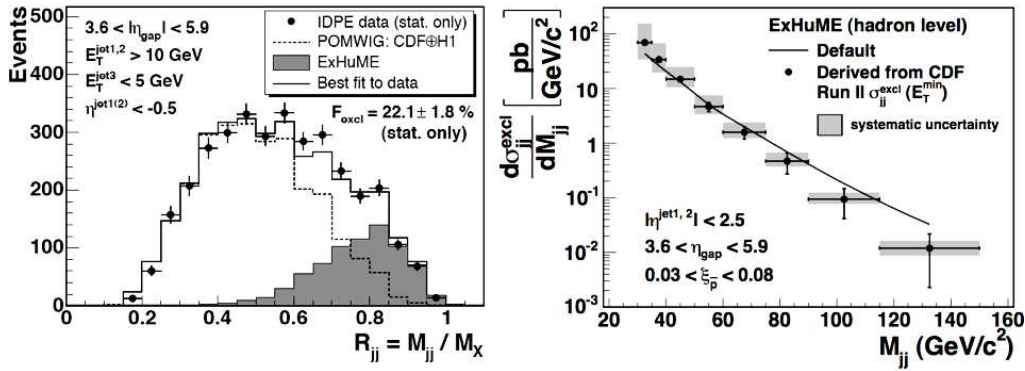}
\caption{(Left) The di-jet mass fraction in DPE data (points) and best fit (solid histogram) to a simulated di-jet mass fraction  obtained
from POMWIG MC events (dashed histogram) and ExHuME di-jet MC events (shaded histogram). 
(Right) The ExHuME~\protect\cite{ExHuME} exclusive di-jet differential cross section at the hadron level vs. 
di-jet mass M$_{jj}$ normalized to measured $\sigma_{jj}^{excl}$ values. 
The curve is the cross section predicted by ExHuME.}
\label{Fig:Exhume-best}
\end{center}
\end{figure}

The process of exclusive di-jet production is important for testing and/or calibrating models
for exclusive Higgs production at the LHC. The CDF~II collaboration has made the first observation of this
process and the main final result is presented in Fig.~\ref{Fig:Exhume-best}.  Details can be found in Ref. 
\cite{PRDDIJET}. This result favours the model of Ref. \cite{KHOZE2000}, which is implemented in the MC simulation
ExHuME \cite{ExHuME}.

\subsection{Exclusive $e^{+}e^{-}$ Production} 

The CDF~II collaboration has reported the first observation of exclusive $e^{+}e^{-}$ 
production in $p\bar p$ collisions~\cite{EXCLEPEM} using 532 pb$^{-1}$ $p\bar{p}$ data collected at $\sqrt{s}= 1.96$ TeV by CDF~II at the Fermilab Tevatron. The definition of exclusivity used  
requires  the absence of any particle  signatures in the detector in the pseudorapidity region $|\eta|<7.4$,
 except for an electron and a positron candidate each with 
transverse energy of $E_{T} \ge 5$ GeV and within the pseudorapidity $|\eta| \le 2$. With these criteria, 
16 events were observed.
%There are four backgrounds to consider:  jets that pass
%electron requirements ($0.0^{+0.1}_{-0.0}$ events), cosmic rays that interact
%in the detector (negligible background), non-exclusive events ($0.3 \pm 0.1$ events), 
%and $\gamma\gamma \rightarrow e^{+}e^{-}$ 
The dominant background is due to events with unobserved  proton dissociation ($1.6 \pm 0.3$ events). The total 
background expectation is  $1.9 \pm 0.3$ events. 
%\begin{figure}[htb]
%\begin{center}
%\includegraphics[width=25pc]{epemkin.eps}
%\caption{A comparison of the LPAIR Monte Carlo prediction for the mass (left) and 
%$\Delta\phi$ (right)  of the exclusive $e^{+}e^{-}$ pair and the data.}
%\label{Fig:epemkin}
%\end{center}
%\end{figure}
%\begin{figure}[htb]
%\begin{center}
%\includegraphics[width=17pc]{epemimnew.eps}
%\caption{A comparison of the LPAIR Monte Carlo prediction for the 
%invariant mass of the exclusive $e^{+}e^{-}$ pair and the data.}
%\label{Fig:epemim}
%\end{center}
%\end{figure}
The observed events are consistent in cross section and properties with the 
QED process $p\bar{p} \rightarrow p + ( e^{+}e^{-}) +  \bar{p}$ through two-photon exchange. 
The measured cross section is $1.6^{+0.5}_{-0.3}(stat) \pm 0.3(syst)$ pb.
This agrees with the theoretical prediction of $1.71 \pm 0.01$ pb obtained using 
the LPAIR MC  generator \cite{LPAIR} and a GEANT based 
detector simulation,  CDFSim \cite{CDFSIM}. 
Details on the observation of the exclusive  $e^{+}e^{-}$ signal are 
reported in reference \cite{EXCLEPEM}.

\subsection{Exclusive $\gamma\gamma$ Production}

An exclusive $\gamma\gamma$ event can be produced via $gg \rightarrow 
\gamma\gamma$ (g = gluon) through a quark loop, with an additional
``screening'' gluon exchanged to cancel the color of the
interacting gluons and so allow the leading hadrons  to stay intact.
This process is closely related \cite{HIGGS, KHOZEGAMGAM} to exclusive
Higgs production at the LHC, $p\bar{p} \rightarrow  p + H + \bar{p}$, where the
production mechanism of the Higgs boson is gg-fusion
through a top quark loop. These processes can also be described as resulting
from DPE.

%\begin{figure}[htb]
%\begin{center}
%\includegraphics[width=17pc]{gamma-gamma-ED1new.eps}
%\caption{An event display of one of the exclusive $\gamma\gamma$ final state 
%candidates.}
%\label{Fig:gammagammaed}
%\end{center}
%\end{figure}

A search has been performed for exclusive $\gamma\gamma$ production in $p$-$\bar{p}$
collisions at $\sqrt{s} = 1.96$TeV, using 532 pb$^{-1}$ of integrated luminosity data  taken by CDF~II at Fermilab. The
event signature requires two electromagnetic showers, each with transverse energy 
$E_{T} \ge  5$ GeV and pseudorapidity $|\eta| \le 1.0$, with no other particles 
detected. Three candidate events were observed. Each candidate  can be
interpreted as either a $\gamma\gamma$ or a  $\pi^{0}\pi^{0}$/$\eta\eta$ 
final state with overlapping photons that  satisfy the $\gamma\gamma$ selection criteria and thus form a background. The probability that  processes 
other than these fluctuate to $\ge 3$ events is $1.7 \times 10^{-4}$. 
%An event display of one of
% the three candidates is given in  Figure \ref{Fig:gammagammaed}.
 %
 %Exclusive $\gamma\gamma$ production was modeled with the ExHume
%Monte Carlo (MC) generator \cite{EXHUME}, based on the calculations of
%references \cite{EXHUMEBASIS1} and \cite{EXHUMEBASIS2}. Simulated single photons, 
%and photons from $\pi^{0}$ and $\eta$ decay, are passed through the GEANT-based
%detector simulation \cite{GEANT} to determine their detection 
%efficiencies.
%
%The 3 candidates are consistent with all being
%$\gamma\gamma$ events but one or more may be $\pi^{0}\pi^{0}$
%or $\eta\eta$. 
Two events clearly favor the $\gamma\gamma$ hypothesis 
and the third event favors the $\pi^{0}\pi^{0}$ hypothesis. On the assumption 
that two of the three candidates are $\gamma\gamma$ events we obtain a cross section
$\sigma(p\bar{p} \rightarrow  p + \gamma\gamma + \bar{p}) = 90^{+120}_{-30} (stat) 
\pm 16(syst)$ fb, for $E_{T} \ge 5$ GeV and $|\eta| \le 1.0$, compatible within the theoretical uncertainties with the
prediction of 40 fb of Ref. \cite{KHOZEAGAIN}. A comparison between the predictions of the ExHuMe MC and the data shows good agreement both in normalization and in the shapes of the kinematic
distributions.
% as indicated in Figure \ref{Fig:gamgamkin}.

%\begin{figure}[htb]
%\begin{center}
%\includegraphics[width=26pc]{gamgamkin.eps}
%\caption{A comparison of data with Monte Carlo (ExHUme)for the 
%invariant mass (left) and $p_{T}$  of the central system.}
%\label{Fig:gamgamkin}
%\end{center}
%\end{figure}

%\begin{figure}[htb]
%\begin{center}
%\includegraphics[width=16pc]{gamgamptnew.eps}
%\caption{A comparison of data with Monte Carlo (ExHUme)for the 
%P_{T}$ of the central system.}
%\label{Fig:gamgampt}
%\end{center}
%\end{figure}

Although two of the candidates are most likely to arise from 
$\gamma\gamma$ production, the $\pi^{0}\pi^{0}$ hypotheses 
cannot be excluded. A 95\% C.L. upper limit is obtained  
on the exclusive $\gamma\gamma$ production cross section ($E_{T}
\ge 5$ GeV, $|\eta| \le 1.0$) of 410 fb, which is about ten
 times higher than the prediction of Ref.~\cite{KHOZEGAMGAM}. This result  
may be used to constrain calculations of exclusive Higgs boson production 
at the LHC. 
Additional CDF data, 
collected with a lower $E_{T}$ threshold,  are being analysed. Exclusive 
$\gamma\gamma$ production has not previously been observed in 
hadron-hadron collisions. This work is described in more detail in 
Ref. \cite{PRLGAMGAM}.

\subsection{Exclusive  $ \mu^{+}\mu^{-} $ Production}

\begin{figure}[htb]
\begin{center}
\includegraphics[width=32pc]{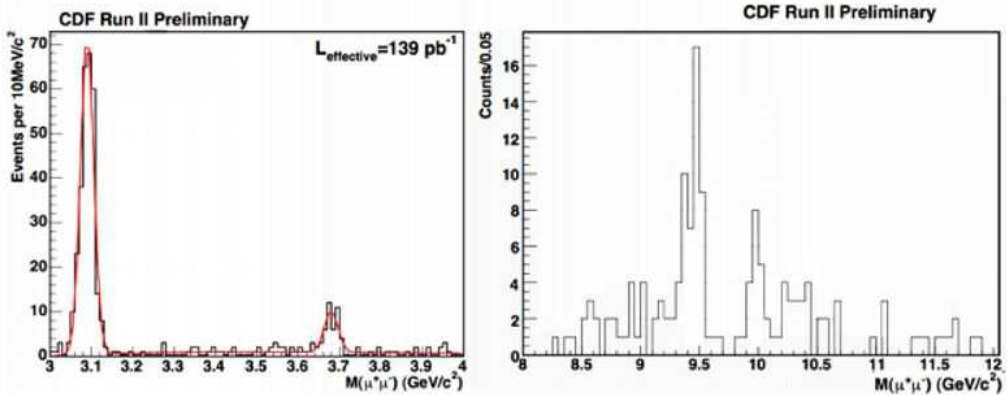}
\caption{(Left) The invariant mass distribution obtained  from the   exclusive $\mu^{+}\mu^{-}$ data;
the $J/\psi$ peak (left) and the smaller $\psi'$ peak (right) can be clearly seen above the
continuum of muon-pair production. (Right) The invariant mass distribution obtained  from the  exclusive higher mass $\mu^{+}\mu^{-}$ data:
the $\Upsilon$ (1S)  peak (middle-left) and the smaller $\Upsilon$ (2S) (middle) peaks can be clearly seen 
above the continuum, while he $\Upsilon $ (3S) peak (middle-right)  is only barely discernible with these statistics.}
\label{Fig:mumuim}
\end{center}
\end{figure}

{\bf{\em Low Mass Exclusive $\mu^{+}\mu^{-}$ Production.}} The CDF~II collaboration has performed a search for exclusive low mass $\mu^{+}\mu^{-}$ final states resulting 
from three processes: $\gamma\gamma\rightarrow$ non-resonant $\mu^{+}\mu^{-}$ 
``continuum'' events, and $J/\psi \rightarrow \mu^{+}\mu^{-}$ \& $\psi'  \rightarrow \mu^{+}\mu^{-}$ events arising from 
$\pom$ - $\gamma$  fusion (photoproduction). In addition, evidence for exclusive $\chi^{\circ}_{c}$ production
was sought arising from the decay channel $\chi^{\circ}_{c} \rightarrow 
J/\psi(\rightarrow \mu^{+}\mu^{-}) + \gamma$. The invariant mass distribution of the
exclusive di-muon events obtained from 1.48 fb$^{-1}$ of data is shown in 
Fig.~\ref{Fig:mumuim}. The $J/\psi$ and $\psi'$ peaks can be clearly seen above
the $\mu^{+}\mu^{-}$ continuum.\footnote{
The offline  cuts applied to the muon-pair data are the same as those applied in the $e^{+}e^{-}$ case: 
there should be no activity in the event in the region $|\eta|<7.4$, and the final state must have two identified muons of $P_{T} >$ 1.4 GeV/c  within $|\eta| < $ 0.6. }
%Cosmic ray contamination was removed from 
%the data sample by requiring:   a  $\Delta$ToF $ < $ 3ns  in the 
%Time of Flight (ToF) counters; and,  tracks that do not have a
%3-D opening angle $\sim$$\pi$ radians i.e.  consistent with with being a single
%straight cosmic  track. 
%\paragraph{Exclusive continuum $ \mu^{+}\mu^{-} $ production}
Continuum $ \mu^{+}\mu^{-} $ production arises from $\gamma\gamma$ interactions. 
These interactions are simulated by the LPAIR  \cite{LPAIR} and STARlight MCs 
\cite{STARLIGHT}. Both give a very good description of the data in  shape 
and in normalization. 
%A comparison 
%of the STARlight Monte Carlo prediction for the exclusive muon pair $\Delta\phi$ 
%and $P_{T}$ compared to the data is shown in Fig.~\ref{Fig:mumukin}.
%\begin{figure}[htb]
%\begin{center}
%\includegraphics[width=25pc]{mumukin.eps}
%\caption{A comparison between the STARLIGHT Monte Carlo prediction for the
%$\Delta\phi$ and $p_{T}$  distribution of the exclusive muon pair for the QED process $\gamma\gamma
%\rightarrow \mu^{+}\mu^{-}$  and the data.}
%\label{Fig:mumukin}
%\end{center}
%\end{figure}
%\begin{figure}[htb]
%\begin{center}
%\includegraphics[width=17pc]{cont_dimuon_pT.eps}
%\caption{A comparison between the STARLIGHT Monte Carlo prediction for the
%$P_{T}$ distribution of the exclusive muon pair and the data.}
%\label{Fig:mumupt}
%\end{center}
%\end{figure}
%\paragraph{Exclusive $ J/\psi (\rightarrow \mu^{+}\mu^{-}) $ and $\psi' 
%(\rightarrow \mu^{+}\mu^{-})$ Photoproduction}
The events  in the $J/\psi$ and $\psi'$ peak of  Figure \ref{Fig:mumuim}, from the process   
$p\bar{p} \rightarrow p + J/\psi(J/\psi^{\prime})+ \bar{p}$, are  mainly
produced via $\pom$ - $\gamma$  fusion. The STARLight MC is used to 
simulate the photoproduction of the $J/\psi$ and the $\psi'$. 
%The 
% odderon\footnote{The odderon is the C-odd partner of the Pomeron - the hard
%odderon skeleton consists of three gluons in a color neutral
%state.  However, as yet, there is no firm evidence for the existence of the odderon}  (O) 
% would contribute to both these final states  via O - $\pom$  fusion. 

%\paragraph{Exclusive $\chi_{c} \rightarrow J/\psi(\rightarrow \mu^{+}\mu^{-})\gamma$ production}

%Previously, a search for the process $p\bar{p} \rightarrow p +  (\chi_{c})  + \bar{p}$, where $\chi_{c} \rightarrow J/\psi(\rightarrow \mu^{+}\mu^{-})\gamma$ was performed at at the Tevatron \cite{wyatt}\footnote{Using 93 pb$^{-1}$ taken by a di-muon trigger, 13 exclusive $J/\psi$ events  and 10 $J/\psi +\gamma$ events were observed with  nothing else visible in the whole detector, and reported as being consistent with originating from $\chi_c$ decays.}.
A $J/\psi$ in the final state can arise from exclusive $\chi^{\circ}_{c}$
production, $p\bar{p} \rightarrow p +  (\chi^{\circ}_{c})  + \bar{p}$ with $\chi^{\circ}_{c} \rightarrow J/\psi (J/\psi \rightarrow \mu^{+}\mu^{-})  + \gamma$.
The photon in 
the $\chi^{\circ}_{c}$ decay  is soft and consequently may not be reconstructed 
and form a  ``background''  to exclusive $J/\psi$ production via $\pom$ - $\gamma$  fusion. The $\chi^{\circ}_{c}$ contributes to the exclusive  $J/\psi$ peak when 
the soft photon from its decay survives the
exclusivity cut.   By fitting the shapes of the E$_{T}$ and $\Delta\phi$ distributions of the di-muon pair of the events in the $J/\psi$ peak of the data with MC generated distributions of $J/\psi$ from photoproduction and $\chi^{\circ}_{c}$ production, CDF~II estimates the $\chi^{\circ}_{c}$ contribution  to the exclusive $J/\psi$ photoproduction peak to be $\approx$ 10\%.

{\bf{\em Higher Mass Exclusive $\mu^{+}\mu^{-}$ Production.}}
The basis of the study of high exclusive muon pairs is somewhat different in that it does not rely on the ``standard'' exclusivity cuts applied to the low mass data.
In this case, one looks for muon pairs that form a vertex with no additional tracks. It is also required that the muons be consistent with $\Delta\phi \approx 0$ and with $P_{T}$-sum approximately zero.
For  890 pb$^{-1}$ of data (2.3M events), with  $\Delta\phi >$ 120$^{o}$ and a $P_{T}$-sum
of the two muon tracks  less than 7 GeV/c, the mass plot shown in 
Fig.~\ref{Fig:mumuim} was obtained.  One can clearly discern the $\Upsilon$(1S) and $\Upsilon$(2) 
peaks in this plot. 
%The $\Upsilon$(3s) is not so convincing, although it is sure to pop 
%out with increased statistics. 
The high mass exclusive muon pair data, with enhanced statistics, 
 is currently under study.

%\begin{figure}[htb]
%\begin{center}
%\includegraphics[width=16pc]{upsilonmass.eps}
%\caption{The invariant mass plot obtained  from the  exclusive higher mass $\mu^{+}\mu^{-}$ data.
%The $\Upsilon$ (1S)  peak (middle-left) and the smaller $\Upsilon$ (2S) peak (middle) can be clearly seen 
%above the continuum.  The $\Upsilon $ (3S) peak (middle-right)  is only barely discernible with these statistics.}
%\label{Fig:upsilon}
%\end{center}
%\end{figure}

\section{Diffractive W/Z Production}
Studies of diffractively produced W/ Z boson are important for understanding the structure of the Pomeron.
The production of intermediate vector bosons is due to the annihllation of quark-antiquark pairs and thus is a  
probe of the quark content of the Pomeron. In leading order, the W/Z is produced by a quark in the
Pomeron, while production by a gluon  is suppressed by a factor of $\alpha_{S}$ and can be distinguished
from quark production by an associated jet \cite{ASSOCJET}. Diffractive dijet production at the Tevatron was found to be suppressed 
by a factor of $\cal{O}$(10) compared  to  expectations from the Diffractive Structure Function (DSF)  extracted
 from diffractive deep inelastic scattering  (DDIS) at the DESY $ep$ Collider HERA.  A more direct comparison 
 could be made by measuring the DSF in diffractive $W$   production at the Tevatron, which is dominated by a 
 $q\bar{q}$ exchange,  as in DDIS. In Run~I, only the overall  diffractive $W$ fraction was measured by CDF~\cite{ASSOCJET}. 
 In Run~II, both the $W$ and $Z$ diffractive fractions and the  DSF are measured.
 
 The CDF Run~II analysis is based on events with RPS tracking from a data sample of $\sim 0.6$~fb$^{-1}$. In addition to the 
 $W/Z$ selection requirements\footnote{
 The CDF $W/Z$ selection requirements are: $E_T^{e,\mu}>25$~GeV, $40<M_T^W<120$~GeV, $66<M^Z<116$~GeV,
 and vertex $z$-coordinate $|z_{vtx}|<60$~cm. 
 In the $W$ case, the requirement of $\xi^{\rm RPS}>\xi^{\rm CAL}$ is very effective in removing the
  overlap evemnts in the region of $\xi^{\rm CAL}<0.1$, while a mass cut of $50<M_W<120$~GeV has the same 
  effect. In the $Z$ case, we use the $\xi^{\rm CAL}$ distribution of all $Z$ events normalized to the RP-track 
  distribution in the region of $-1<\log\xi^{\rm CAL}<-0.4$ ($0.1<\xi^{\rm CAL}<0.4$) to obtain the ND background 
  in the diffractive region of $\xi^{\rm CAL}<0.1$.},   
  a hit in the RPS trigger counters and a RPS reconstructed  track with $0.03<\xi<0.1$ and $|t|<1$ are required. 
A novel feature of the analysis is the determination of the full kinematics of the $W\rightarrow e\nu/\mu\nu$ 
decay using the neutrino $E_T^\nu$ obtained from the missing $E_T$, as usual, and $\eta_\nu$ from the formula 
$\xi^{\rm RPS}-\xi^{\rm cal}=(E_T/\sqrt{s})\exp[-\eta_\nu]$ , where $\xi^{\rm cal}=\sum_{\rm towers}(E_T/\sqrt{s})\exp[-\eta]$. 
%The $W$ mass distribution for events with $\xi^{\rm CAL}<\xi^{RPS}$ is 
 %shown in Fig.~\ref{Fig:W-mass} along with a Gaussian fit. 
 The extracted value of $M_W^{\rm exp}=80.9\pm 0.7$~GeV 
 is in good agreement with the world average $W$ mass of $M_W^{\rm PDG}=80.403\pm 0.029$~GeV~\cite{PDG}. 
After applying corrections ccounting for the RPS acceptance, $A_{\rm RPS}\approx 80$~\%, the trigger counter efficiency, $\epsilon_{\rm RPStrig}\approx 75$~\%, the track reconstruction efficiency, $\epsilon_{\rm RPStrk}\approx 87$~\%, multiplying by 2 to include production by $p\bar{p} \rightarrow W/Z+p$, and correcting the ND event number for the effect of overlaps due to multiple interactions by multiplying by the factor $f_{\rm 1-int} \approx 0.25$,
%\footnote{Accounting for the RPS acceptance $A_{\rm RPS}\approx 80$~\%, the trigger counter efficiency $\epsilon_{\rm RPStrig}\approx 75$~\%, the track reconstruction efficiency $\epsilon_{\rm RPStrk}\approx 87$~\%, multiplying by 2 to include production by $p\bar{p} \rightarrow W/Z+p$, and correcting the ND event number for the effect of overlaps due to multiple interactions by multiplying by the factor $f_{\rm 1-int} \approx 0.25$.}
the diffractive fraction of $W/Z$ events was obtained as
 $R_{W/Z}=2\cdot N_{SD}/A_{\rm RPS}/\epsilon_{\rm RPStrig}/\epsilon_{\rm RPStrk}/(N_{\rm ND}\cdot f_{\rm 1-int})$: 
 \begin{eqnarray}
 & R_W(0.03 < \xi< 0.10,\,|t|<0.1)=[0.97\pm 0.05\;\mbox{(stat)}\pm 0.11\;\mbox{(syst)]}\% \\
& R_Z(0.03<\xi<0.10,\,|t|<0.1)=[0.85\pm 0.20\;\mbox{(stat)}\pm 0.11\;\mbox{(syst)]}\%
\end{eqnarray}
The $R_W$ value is consistent with the Run~I result of $R_W(0.03<\xi<0.10,\,|t|<0.1)=[0.97\pm 0.47]$ obtained from the published value of $R^W(\xi<0.1)=[0.15\pm0.51\;\mbox{(stat)}\pm0.20\;\mbox{(syst)}]\%$~\cite{ASSOCJET}  multiplied by a factor of 0.85 that accounts for the reduced ($\xi$-$t$) range in Run~II.
\section{Conclusion}
We present recent results on exclusive central prodction of di-jets, di-leptons, and di-photons reported by the CDF~II collaboration, obtained from Run~II data collected at the Tevatron $p\bar p$ collider at $\sqrt s=1.96$~TeV. The results are compared with theoretical expectations, and implications for the possible observation of exclusive Higgs boson production and other interesting new physics processes at the Large Hadron Collider are discussed.       
%------------------------------------------------------------------------------
%       Bibliography
%------------------------------------------------------------------------------
\bibliographystyle{heralhc} 
{\raggedright
\bibliography{goulianos-pinfold}
}

\end{document}